\documentclass{pasj00}
\draft

\begin{document}
\SetRunningHead{Takagi et al.}{Disk Dissipation Timescale in Taurus}
%\Received{}%{yyyy/mm/dd}
%\Accepted{}%{yyyy/mm/dd}
%\Published{}%{yyyy/mm/dd}

\title{Disk Dissipation Timescale of Pre-main Sequence Stars in Taurus\thanks{Based in part on data collected at Subaru Telescope, which is operated by the National Astronomical Observatory of Japan.}}

%%% begin:list of authors
% Do NOT capitalize all letters in "textsc".
\author{Yuhei \textsc{Takagi} and Yoichi \textsc{Itoh}}
\affil{Nishi-Harima Astronomical Observatory, Center of Astronomy, University of Hyogo, 407-2, Nishigaichi, Sayo, Sayo, Hyogo 679-5313}
\email{takagi@nhao.jp}
\and
\author{Yumiko {\sc Oasa}}
\affil{Faculty of Education, Saitama University, 255 Shimo-Okubo, Sakura, Saitama, Saitama 388-8570}
%%% end:list of authors

%%% Please use the following style in case that sorting by 
%%% affiliation is impossible. 
%
% \author{%
%   D-Firstname \textsc{D-Familyname}\altaffilmark{1}
%   E-Firstname \textsc{E-Familyname}\altaffilmark{1,2}
%   and
%   F-Firstname \textsc{F-Familyname}\altaffilmark{2}}
% \altaffiltext{1}{Address of Institute}
% \email{ddddd@xxx.xxx.xx.xx}
% \email{eeeee@xxx.xxx.xx.xx}
% \altaffiltext{2}{Address of Institute}

%% `\KeyWords{}' always has to be placed before `\maketitle'.
 %Do NOT move this preamble from here!

\maketitle

\begin{abstract}
We present the results of an age determination study of pre-main sequence stars in the Taurus star-forming region. The ages of 10 single stars with masses of 0.5---1.1 $M_{\odot}$ were derived from the surface gravities, estimated from high-resolution optical and near-infrared spectroscopy. The equivalent width ratios of nearby absorption line pairs were employed for the surface gravity diagnostic, which directly reflects the parameters of the stellar atmosphere without any veiling correction. From a comparison of determined ages and near-infrared color excesses such as $J-H$, $J-K$, and $J-L$, the inner disk lifetime of the young stars with 0.5---1.1 $M_{\odot}$ in Taurus is estimated to be 3---4 Myr.
\end{abstract}

\section{Introduction}

Determining the ages of pre-main sequence (PMS) stars is critical to understanding the formation and evolution of stars, protoplanetary disks, and planets. Recent direct imaging observations have discovered various evolutionary states of the disks and the planets (e.g., \cite{Hashimoto2012}, \cite{Tanii2012}). Precise age determinations of these PMS stars will make a significant contribution to the understanding of the evolution of young stars.

The ages of young stars are mainly estimated by comparing their loci and the theoretical evolutionary tracks on the Hertzsprung-Russell (HR) diagram, using luminosities and effective temperatures obtained from broadband photometry (e.g., \cite{Strom1989}, \cite{Kenyon1995}). The members of a single star-forming region do not fall along a specific isochrone, indicating an age spread among the members. Moreover, the plots of Class II and Class III stars often seem to be mixed on the HR diagram, which makes it difficult to determine the evolutionary timescale of disks and planets. 

Before discussing the PMS star evolution, we need to verify the estimated ages. Accurate luminosities of PMS stars are difficult to estimate due to uncertainties in distance, extinction, and excesses. The excesses are seen in both the ultraviolet and the near- to far-infrared wavelengths, which arise from the accretion shock on the surface of the photosphere and the heated disk, respectively. Many studies have attempted to determine stellar luminosities with high precision. \citet{Bertout2007} estimated the ages of the members in the Taurus-Auriga star-forming region based on the distance estimates from a kinematic study \citep{Bertout2006}. The luminosities of the members were calculated by combining the derived distances with the $I$-band flux, of which the excess from the circumstellar material is considered to be small \citep{Cieza2005}. Using the recalculated luminosities, the loci of the classical T Tauri stars and weak-lined T Tauri stars were separated, and the disk lifetime in Taurus-Auriga was estimated to be 4.0 Myr for solar mass stars. In addition, many studies have characterized the excesses in both ultraviolet and near-infrared radiation in order to determine actual stellar parameters (e.g., \cite{Fischer2011}, \cite{McClure2013}).

Instead of determining the luminosities from photometric studies, surface gravity $g$ (cm/$\mathrm{s}^2$) measurements of young stars allow age estimations, because its photosphere is still contracting. The surface gravity of a PMS star can be estimated by measuring the strength of gravity sensitive absorption features. \citet{Steele1995} used the Na doublet in the $I$-band obtained with low resolution spectroscopy to distinguish the young members in the Pleiades cluster from field stars. \citet{Slesnick2006} also applied these absorption lines as a gravity indicator in order to verify the luminosity class of late type stars in the Upper Scorpius OB Association. \citet{Schiavon1995} compared the observed CaH molecular lines of young stars in the Taurus star-forming region with the theoretical spectra for gravity estimation. 

To discuss the evolution timescale of PMS stars, disks, and planets from the age estimation based on the spectroscopic method, an accurate gravity estimation is necessary. In our previous studies, we established the surface gravity indicators based on high resolution spectroscopy. We used the equivalent width ratios (EWRs) of the Fe (8186.7$\AA$, 8204.9$\AA$) and Na (8183.3$\AA$, 8194.8$\AA$) absorption lines as surface gravity indicators for late-K type PMS stars \citep[hereafter Paper I]{Takagi2010}. When veiling is present, absorption lines are partially filled in, but this effect was removed by calculating the EWRs of nearby absorption lines, as the veiling effect is nearly uniform over such a limited wavelength range. The EWRs of Sc lines (22057.8$\AA$ and 22071.3$\AA$) and Na lines (22062.4$\AA$ and 22089.7$\AA$) have also been established as a gravity diagnostic for stars with spectral types of late-K to early-M \citep[hereafter Paper II]{Takagi2011}. 

In Papers I and II, the ages of five PMS stars in Taurus were estimated, providing the age estimations based on methods independent of distance, extinction, and veiling. In the present paper, the evolutionary timescale of PMS stars with 0.5---1.1 $M_{\odot}$ in the Taurus star-forming region is presented by combining the results of these previous studies with ages of the PMS stars obtained from additional observations. In Section 2, new observations of Class II stars and transitional disk objects are described. In Section 3, the results of Paper I, Paper II, and the new observations are summarized with error considerations. The disk lifetime of PMS stars in Taurus is discussed in Section 4, based on the relationship between the estimated ages and other parameters such as near-infrared color excesses and emission lines of H$\alpha$ and [O\emissiontype{I}] 6300$\AA$.

\section{Observations and Data Reductions}

\subsection{Optical Spectroscopy}

High-resolution spectroscopic observations of two Class II objects and two transitional disk objects were conducted on 7 December 2010 with the High Dispersion Spectrograph (HDS) mounted on the Subaru Telescope. In order to use the EWR - $g$ relationships derived in Paper I, late-K type PMS stars were selected for the targets. The slit width was set to 0".6 to achieve a resolution of $\sim$60000. The settings of the optical elements were optimized to obtain the Fe (8204.9$\AA$) and Na (8183.3$\AA$, 8194.8$\AA$) lines in the 73$^{\mathrm {rd}}$ order of the echelle spectrograph. The integration time for each object was set to 1800s to achieve the signal-to-noise ratio (S/N) of 100. The spectra of A0 type stars were obtained as standard stars for telluric absorption correction. Th-Ar lamp frames were taken at the beginning and the end of the observation for wavelength calibration.

Data reduction was performed with the Image Reduction and Analysis Facility (IRAF) software package\footnote{IRAF is distributed by the National Optical Astronomy Observatory.}. Bias subtraction, flat fielding, cosmic ray rejection, and scattered light subtraction were first conducted. Then the spectra were extracted using the APALL task. The wavelength was calibrated with the Th-Ar comparison spectrum. After removing the telluric absorption lines and performing continuum normalization, the resultant spectra were obtained. The SPLOT task was used to measure the equivalent widths of the absorption lines. The EWRs using the Fe line at 8186.7$\AA$ represented in Paper I were not employed in this study due to the heavily blended telluric line. The Voigt function was adopted for line fitting. The errors in equivalent widths were calculated from the uncertainties in the continuum level. The equivalent widths were measured with an identical method. The measured equivalent widths of four stars and the equivalent widths of two stars mentioned in Paper I are listed in table \ref{table-ew_op}.

\subsection{Near-Infrared Archival Data}

The high-resolution spectra of PMS stars observed in the $K$-band were searched for in the Keck Observatory Archive. A spectrum of CY Tau taken with NIRSPEC on 11 February 2006 satisfied the requirement of this research. Since the spectral type of CY Tau is K7, the EWR - $g$ relationships estimated in Paper II is suitable for estimating its gravity and age. The resolution power of this observation was $\sim$19000 with the 0".576$\times$12" slit. To subtract the sky emissions, the data were obtained with an ABBA nodding pattern. The total integration time was 520s, resulting in an S/N of 170. The spectrum of an A0 type star (HR 1237) was obtained for the telluric line correction.

IRAF was employed again for the reduction of near-infrared data. Sky and dark currents were removed by subtracting the frame obtained in the opposite nod position, and then flat fielding, cosmic ray rejection, and aperture extraction were conducted. Wavelength calibration and distortion correction were carried out simultaneously using OH emission lines. After the spectra were extracted with APALL, telluric line correction, spectral combining, and continuum normalization were executed. During the equivalent width measurements, Gaussian and Voigt functions were used for the Sc and Na line fits, respectively. In Paper II, EWRs that included the Sc line at 22071.3$\AA$ were established as surface gravity indicators, however, this line was not used in this study because of the heavy blending with the Si line at 22068.7$\AA$. The equivalent widths of CY Tau and the three stars mentioned in Paper II measured with a uniform method are presented in table \ref{table-ew_nir}.

\section{Results}

\subsection{Surface gravity estimations}

\begin{figure}
 \begin{center}
  \includegraphics[width=16cm]{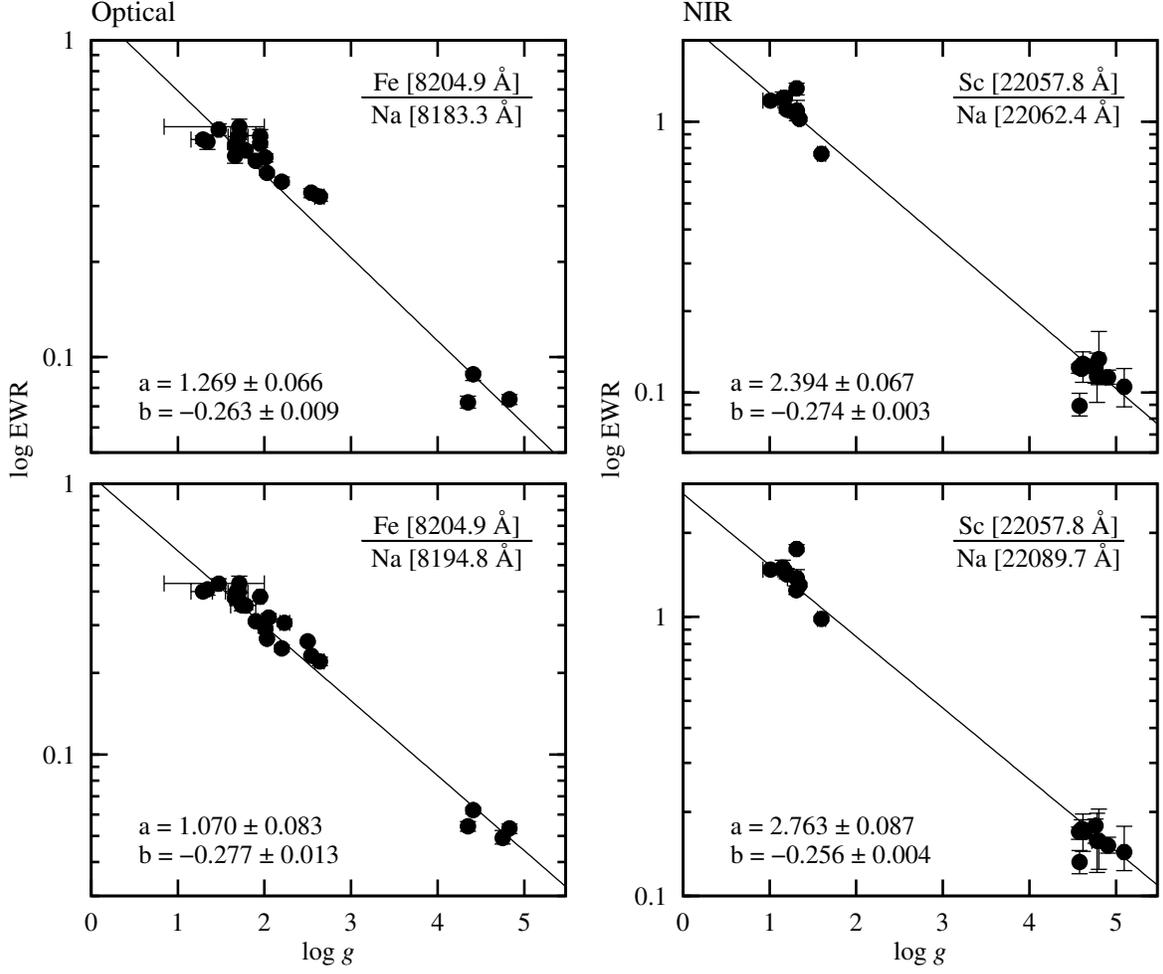} 
 \end{center}
\caption{The relations between the EWRs and the surface gravities. Left two panels show the EWR - $g$ relations using the Fe and Na lines in the $I$-band. Right two panels show those of NIR lines. The solid line in each panel is the approximate curve expressed as EWR $ = ag^b$. The values of $a$ and $b$ for each relationship is also presented.}\label{EWR}
\end{figure}

The surface gravities of the PMS stars can be determined by comparing the observed EWR with the EWR - $g$ relationships described in Papers I and II. Before executing the surface gravity estimations, we revised the EWR - $g$ relationships. As mentioned in Papers I and II, these relations are derived from the observations of field dwarfs and giants of which the parallaxes, spectral types, and the $V$-band magnitudes are well studied in the previous works. We used the newly analyzed Hipparcos parallaxes \citep{vanLeeuwen2007} to all the objects included in these papers. The magnitudes of the objects were quoted from Tycho-2 catalog \citep{Hog2000}. The mass of each giant star was recalculated by comparing the location in the HR diagram and the theoretical evolutional tracks \citep{Lejeune2001}, as mentioned in Paper I. Giant stars whose masses were calculated to be smaller than 1 $M_{\odot}$ were removed since the ages of these stars are estimated to be beyond the age of the universe. The revised EWR - $g$ relationships are shown in figure \ref{EWR}. We conducted a power-law fitting in each relationship using the equation given by EWR $ = ag^b$, since the equivalent width of the atomic absorption lines can be approximated by a power of the surface gravity (e.g., \cite{Gray2005}). The values and the standard deviations of $a$ and $b$ in each relation estimated by the weighted least square fit are presented in figure \ref{EWR}.

Using the estimated EWR -log $g$ relationship, the surface gravities of the PMS stars and the errors are estimated with the equations in the Appendix. Table \ref{table-age} shows the derived surface gravities of the 5 PMS stars addressed in Papers I and II and the 5 PMS stars observed in the present work. The major cause of the error is the uncertainties of the equivalent widths which are attributed to the S/N of the observed spectra. The typical error of log $g$ arising from this part is 0.1 for the spectrum with S/N of 100$\sim$120. The uncertainties of the optical and NIR EWR - $g$ relationships cause the log $g$ errors of 0.09$\sim$0.13 and 0.05, respectively, within a range of log $g$=3.5 to 4.0. 

\subsection{Age estimations}

The ages were calculated by comparing the surface gravities and the effective temperatures to the evolutionary model of the PMS stars \citep{Siess2000}. Effective temperatures of the objects were estimated from the spectral types derived from photometric researches (\cite{Kenyon1995}, \cite{Furlan2006}). Masses were calculated from their loci on the HR diagram. The errors in ages were estimated from the uncertainty in the surface gravity for each object. The derived properties of the targets are listed in table \ref{table-age}. Since the effective temperatures of the objects are estimated from the spectral types obtained from the photometric methods with no errors, the uncertainty of the effective temperature is not included in the estimated age. The spectral type of CY tau is indicated as K7 in \citet{Furlan2006}, whereas \citet{Kenyon1995} designated it as M1. This discrepancy makes a 0.5 Myr difference in the age. In order to estimate the age and the effective temperature of the PMS stars with higher accuracy, the EWR method using the temperature sensitive lines is expected.

\begin{figure}
 \begin{center}
  \includegraphics[width=8cm]{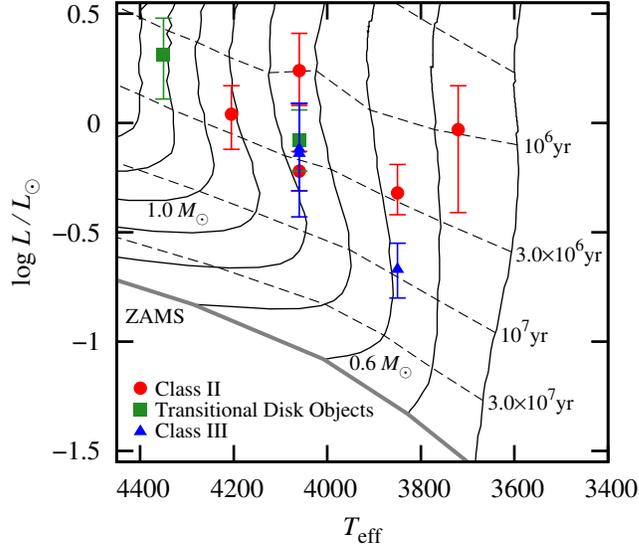} 
 \end{center}
\caption{The HR diagram of the observed stars. Red circles, green squares, and blue triangles denote Class II objects, transitional disk objects, and Class III objects, respectively. The solid lines are evolutionary tracks (Siess et al. 2000) for masses of 0.4 to 1.2 $M_{\odot}$. The dashed lines indicate isochrones.}\label{HRD}
\end{figure}

The HR diagram of the targets is presented in figure \ref{HRD}. The weighted average age of the Class II and transitional disk objects is 1.4$\pm$0.5 Myr, while the average age of Class III is 3.5$\pm$1.6 Myr. We carried out the two-sample Kolmogorov-Smirnov (K-S) test in order to estimate the probability that these two samples are drawn from different distributions. The calculated p-value was 0.17 which shows this probability cannot be rejected with 0.05 significance level. We added a pseudo data to each group in order to estimate the number of additional objects needed to distinguish these two groups with a reliable probability. These data were obtained with gaussian random number generation based on Box-Muller method. As a result, two groups can be discriminated with $<$ 0.05 significance level by adding $\sim$5 objects to either group or adding 3$\sim$4 objects to both groups. The trend that the averaged age of Class II + transitional disk objects is younger than that of Class III is consistent with \citet{Bertout2007}, while the individual age of each star is inconsistent. The cause of this difference is may probably be due to the uncertainty in distance, extinction, and veiling. Since the EWRs are free from these contaminants, our age determination may be reliable, though the uncertainties exist on the surface gravities of the giants and the dwarfs in EWR - $g$ relationship estimation due to the parallax errors, and the effective temperatures of the pre-main sequence stars derived from the SED fittings.  Because age determination depends on the choice of evolutionary model for these young stars \citep{Hillenbrand2004}, the ages of the target stars were also estimated following \citet{Baraffe1998}. Although the difference in masses adopted between the models was 20\% at most, the calculated ages were comparable.

\section{Discussions}

By combining the estimated ages of PMS stars with many observational results, such as determinations of the spectral energy distribution (SED) and the mass accretion rate, the lifetime of the disk surrounding the PMS stars was deduced to be a few Myr. \citet{Haisch2001} derived disk lifetime by comparing the ages of young star clusters and the fraction of stars with circumstellar disks. They used the $JHKL$ colors to estimate the disk fraction in each star-forming region, and found that half of PMS stars lose their disks in 3 Myr. However, this trend is still unproven within a single star-forming region. Because compositional differences in the initial conditions of a molecular cloud (such as mass, metallicity, and density) may lead to variation in the evolution of stars and disks, the disk dissipation timescale should be discussed in a single environment. 

To derive the disk evolution timescale in Taurus, the estimated ages with EWRs were compared to the near-infrared color excesses of each object. Near-infrared excess will decrease with increasing age if the disk dissipates with time. Near-infrared magnitudes of the objects were taken from \citet{Kenyon1995}, and then the intrinsic near-infrared color excesses were calculated by subtracting the photospheric color and adopting the reddening law (\cite{Cohen1981}, \cite{Meyer1997}) using the $A_V$ values from \citet{Kenyon1995} and \citet{Furlan2006}. The color excesses are listed in table \ref{table-color}. Two transitional disk objects were included in the targets. LkCa 15 is believed to be a pre-transitional object, in which a gap exists between the small inner disk (0.12-0.15 AU) and the outer disk beyond 46 AU \citep{Espaillat2007}. V819 Tau is classified as Class III in \citet{Furlan2006}, however, a small excess exists in wavelengths longer than 10 $\mu$m. Therefore, it could be considered as a star in a later transitional phase.

\begin{figure}
 \begin{center}
  \includegraphics[width=16cm]{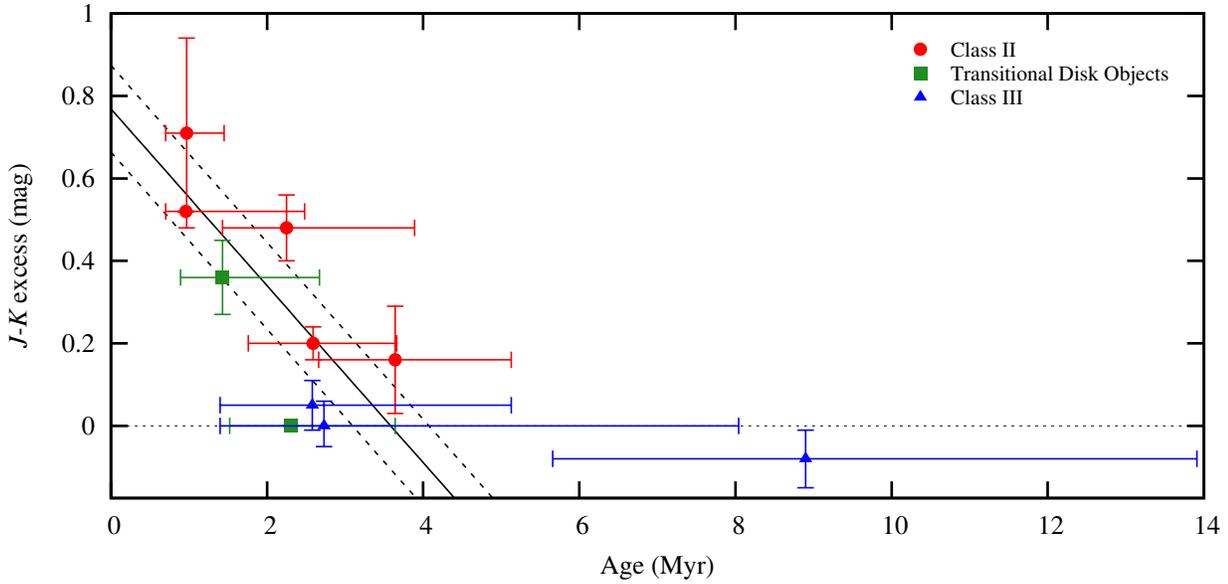} 
 \end{center}
\caption{Age - $J-K$ excess relationship of the observed stars. Red circles, green squares, and blue triangles represent Class II objects, transitional disk objects, and Class III objects, respectively. The solid line indicates the regression line derived from the plots with the $J-K$ excess $>$ 0. The dashed lines are drawn based on the 1 $\sigma$ uncertainty of the regression line at $J-K$ excess = 0.}\label{age-color}
\end{figure}

Figure \ref{age-color} shows the relationship between the age and the $J-K$ excess. This is the first observational evidence which represents disk dissipation with an increment in age within a single star-forming region. Near-infrared excesses decrease at a constant rate, indicating that the excess does not vanish on a short timescale. This trend was also seen when the estimated ages were compared to the $J-H$ and $J-L$ excesses. The regression lines were estimated by the weighted least square fitting, using plots with colors greater than zero in each age - color excess relation. The excess dissipation timescales in $J-H$, $J-K$, and $J-L$ were 2.6$\pm$1.4 Myr, 3.6$\pm$0.5 Myr, and 4.4$\pm$1.6 Myr, respectively. The error in each timescale is calculated from the standard deviations of the regression line coefficients. 

The $JHKL$ excesses indicate the presence of an inner disk. Many mechanisms for creating an inner hole have been suggested by theoretical studies; they include photoevaporation (e.g., \cite{Clarke2001}, \cite{Alexander2006}), grain growth (e.g., \cite{Tanaka2005}), and planet formation (e.g., \cite{Zhu2011}). In these studies, disk lifetime was estimated to be a few Myr to 10 Myr, which is consistent with our results and other observations (e.g., Haisch et al. 2001, Bertout et al. 2007). However, the linear relationship between age and $JHKL$ excesses represented in the present study suggests slow dissipation of the inner disk. This result contradicts the outcome suggested by theoretical photoevaporation studies, which have indicated that the timescale for inner hole creation and disk dissipation is $\sim10^5$yr \citep{Alexander2006}. 

We also assessed the relationship between age and the equivalent widths of the H$\alpha$ line and the [O\emissiontype{I}] 6300$\AA$ line in order to estimate dissipation timescales of the accretion and the disk wind. The equivalent width of the H$\alpha$ line and the [O\emissiontype{I}] line were quoted from \citet{Strom1989} and \citet{Hartigan1995}, respectively. The equivalent width of [O\emissiontype{I}] line in LkCa 15 was taken from \citet{Iguchi2014}. The data was obtained from the Keck Observatory Archive where the resolution power was 50000 and the exposure time was 1200s. After the bias subtraction, flat fielding, spectra extraction and the wavelength calibration, the veiling correction was conducted by fitting the absorption lines to a dwarf spectrum. All the equivalent widths of these lines are listed in table \ref{table-color}.

\begin{figure}
 \begin{center}
  \includegraphics[width=16cm]{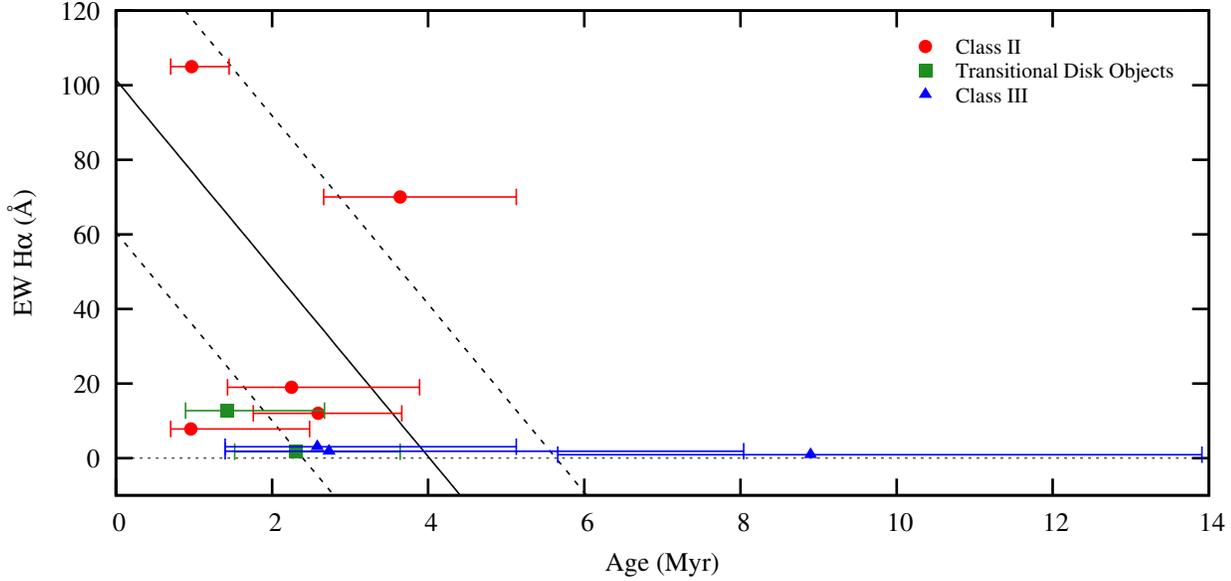} 
 \end{center}
\caption{Age - H$\alpha$ equivalent width relationship for the observed stars. Red circles, green squares, and blue triangles indicate Class II objects, transitional disk objects, and Class III objects, respectively. The solid line represents the regression line. The dashed lines are drawn based on the 1 $\sigma$ uncertainty of the regression line at EW H$\alpha$ = 0.}\label{age-Ha}
\end{figure}

\begin{figure}
 \begin{center}
  \includegraphics[width=16cm]{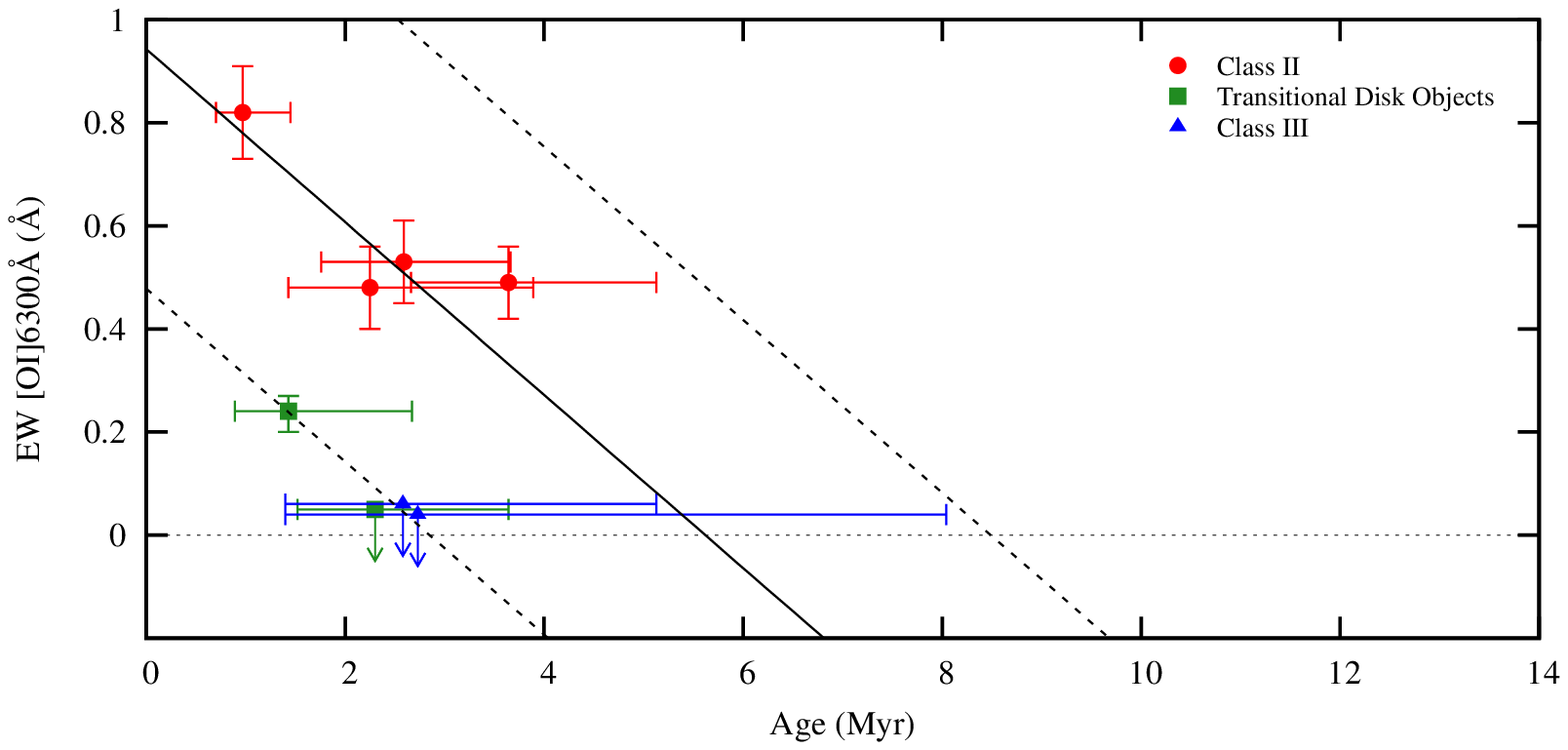} 
 \end{center}
\caption{Age - [O\emissiontype{I}] equivalent width relationship for the observed stars. Red circles, green squares, and blue triangles indicate Class II objects, transitional disk objects, and Class III objects, respectively. The plots with arrow indicate the upper limit. The solid line represents the regression line. The dashed lines are drawn based on the 1 $\sigma$ uncertainty of the regression line at EW [O\emissiontype{I}] = 0.}\label{age-OI}
\end{figure}

The estimated dissipation timescales of the H$\alpha$ and the [O\emissiontype{I}] emissions are 4.0$\pm$1.6 Myr (figure \ref{age-Ha}) and 5.6$\pm$2.8 Myr (figure \ref{age-OI}), respectively. These timescales are comparable to those of the infrared excesses though a large uncertainty exists. Nevertheless, it was difficult to determine whether the disk accretion and the wind are decreasing drastically or modestly. A correlation between the equivalent width of H$\alpha$ line and the inclination angle is found (e.g., \cite{Appenzeller2005}, \cite{Kurosawa2006}). The dispersion in figure \ref{age-Ha} may be reduced by correcting the inclination. In addition, \citet{Appenzeller2013} suggested the inclination dependency on the wind velocity derived from the blue edge of the [O\emissiontype{I}] line. Comparison of the ages of PMS stars and the inclination corrected wind velocity derived from the [O\emissiontype{I}] line may be practical for discussing disk evolution. However, because the inclinations are known for only four of our objects, the inclination correction was not conducted. By observing more PMS stars with known inclination or determining the inclination of the targets in the present work may improve the relationship between the age and equivalent widths of the H$\alpha$ and [O\emissiontype{I}] lines. Also, obtaining the equivalent width of the [O\emissiontype{I}] in stars with optically thin disks may improve the estimation of the [O\emissiontype{I}] dissipation timescale. We conclude that the lifetime of the inner disk of PMS stars in Taurus with masses of 0.4---1.1 $M_{\odot}$ is 3.0---4.0 Myr, based on the age - color excess and the age - emission relations.

\bigskip

This research was based on data collected at the Subaru Telescope, which is operated by the National Astronomical Observatory of Japan. This research has made use of the Keck Observatory Archive (KOA), which is operated by the W. M. Keck Observatory and the NASA Exoplanet Science Institute (NExScI), under contract with the National Aeronautics and Space Administration.

\appendix
\section*{Error of the Surface Gravity Calculation}
Surface gravity of a PMS star can be calculated from the following equation due to the EWR - $g$ relationship,
\begin{equation}
  \log g = \frac{\log EWR - \log a}{b}. 
\end{equation}
The error of log $g$ is calculated by using the error propagation,
\begin{equation}
  \delta \log g = \sqrt{\left(-\frac{1}{ab\ln 10}\delta a\right)^2+\left(\frac{\log a - \log EWR}{b^2}\delta b\right)^2+\left(\frac{1}{bEWR\ln 10}\delta EWR\right)^2}. 
\end{equation} 

\clearpage

%%%%%%%%%%%%%%%%%%%%%%%%%%%%%%%%%%%%%%%

%%% Fig. 1 %%%

%%% Fig. 2 %%%

%%% Fig. 3 %%%

%%% Fig. 4 %%%

%%% Fig. 5 %%%

\clearpage

%%%%%%%%%%%%%%%%%%%%%%%%%%%%%%%%%%%%%%%%

\begin{table}
 \begin{center}
  \caption{Measured equivalent widths of stars observed in the optical $I$-band.}\label{table-ew_op}
   \begin{tabular}{lccc}
   \hline \hline
   Star 			& EW(Na 8183.3$\AA$)		& EW(Na 8194.8$\AA$)		& EW(Fe 8204.9$\AA$)		\\ 
			& ($\AA$)				& ($\AA$)				& ($\AA$)	 			\\ \hline
   DL Tau			& 0.326$^{+0.018}_{-0.016}$		& 0.350$^{+0.017}_{-0.024}$		& 0.045$^{+0.004}_{-0.005}$		\\ 
   GI Tau			& 0.731$^{+0.023}_{-0.023}$		& 0.913$^{+0.005}_{-0.057}$		& 0.086$^{+0.007}_{-0.007}$		\\ 
   LkCa 15		& 0.528$^{+0.027}_{-0.029}$		& 0.692$^{+0.037}_{-0.034}$		& 0.069$^{+0.009}_{-0.007}$		\\ 
   V819 Tau		& 0.817$^{+0.012}_{-0.027}$		& 0.974$^{+0.030}_{-0.022}$		& 0.095$^{+0.007}_{-0.008}$		\\ 
   HBC 374		& ---				& 0.956$^{+0.044}_{-0.019}$		& 0.086$^{+0.007}_{-0.007}$		\\ 
   V827 Tau		& ---				& 1.271$^{+0.049}_{-0.062}$		& 0.112$^{+0.018}_{-0.012}$		\\ 
   \hline
  \end{tabular}
 \end{center}
\end{table}

\begin{table}
 \begin{center}  
  \caption{Measured equivalent widths of stars observed in the near-infrared $K$-band.}\label{table-ew_nir}
   \begin{tabular}{lccc}
   \hline \hline
   Star 		& EW(Sc 22057.8$\AA$)		& EW(Na 22062.4$\AA$)		& EW(Na 22089.7$\AA$)		\\ 
		& ($\AA$)				& ($\AA$)				& ($\AA$)	 			\\ \hline
   CY Tau		& 0.214$^{+0.008}_{-0.001}$		& 1.090$^{+0.100}_{-0.077}$		& 0.833$^{+0.077}_{-0.059}$		\\ 
   DN Tau		& 0.331$^{+0.008}_{-0.002}$		& 1.580$^{+0.133}_{-0.168}$		& 1.179$^{+0.112}_{-0.119}$		\\
   IQ Tau		& 0.216$^{+0.030}_{-0.026}$		& 0.777$^{+0.255}_{-0.093}$		& 0.598$^{+0.196}_{-0.072}$		\\
   LkCa 14	& 0.243$^{+0.005}_{-0.004}$		& 1.447$^{+0.162}_{-0.166}$		& 1.105$^{+0.123}_{-0.128}$		\\
   \hline
   \end{tabular}
  \end{center}
\end{table}

\begin{table}
 \begin{center} 
  \caption{\bf{Properties of the observed PMS stars.}}\label{table-age}
   \begin{tabular}{lcccccccc}
   \hline \hline
   Star 		& Type\footnotemark[$*$] 	& SpT\footnotemark[$\dagger$]		& $T_{\mathrm{eff}}$\footnotemark[$\ddagger$] 	& Method\footnotemark[$\S$]	& log $g$					& $M_{\rm star}$\footnotemark[$\|$]			& Age						& log $L_{\rm star}/L_{\odot}$ 	\\ 
				&						&						& (K)						&						& 							& ($M_{\odot}$) 			& (Myr) 					& 				 	\\ \hline
CY Tau 	& II	& K7$^a$	& 4060	& NIR	& 4.00$^{+0.11}_{-0.10}$	& 0.78$^{+0.02}_{-0.02}$	& 3.64$^{+1.49}_{-0.98}$	& -0.22$^{+0.09}_{-0.09}$	\\ 
DL Tau 	& II	& K7$^a$	& 4060	& Optical	& 3.51$^{+0.17}_{-0.17}$	& 0.74$^{+0.01}_{-0.00}$	& 0.97$^{+0.48}_{-0.27}$	& 0.24$^{+0.17}_{-0.16}$	\\ 
DN Tau 	& II	& M0$^a$	& 3850	& NIR	& 3.87$^{+0.11}_{-0.13}$	& 0.57$^{+0.01}_{-0.00}$	& 2.59$^{+1.07}_{-0.83}$	& -0.32$^{+0.13}_{-0.10}$	\\ 
GI Tau 	& II	& K6$^a$	& 4205	& Optical	& 3.87$^{+0.17}_{-0.15}$	& 0.92$^{+0.02}_{-0.02}$	& 2.25$^{+1.64}_{-0.82}$	& 0.04$^{+0.13}_{-0.16}$	\\ 
IQ Tau 	& II	& M1$^a$	& 3720	& NIR	& 3.43$^{+0.40}_{-0.20}$	& 0.47$^{+0.01}_{-0.00}$	& 0.96$^{+1.52}_{-0.26}$	& -0.03$^{+0.20}_{-0.38}$	\\ 
LkCa 15 	& TD	& K5$^a$	& 4350	& Optical	& 3.74$^{+0.20}_{-0.17}$	& 1.11$^{+0.01}_{-0.00}$	& 1.43$^{+1.24}_{-0.54}$	& 0.31$^{+0.17}_{-0.20}$	\\ 
V819 Tau 	& TD	& K7$^a$	& 4060	& Optical	& 3.85$^{+0.15}_{-0.15}$	& 0.76$^{+0.02}_{-0.01}$	& 2.30$^{+1.34}_{-0.78}$	& -0.08$^{+0.14}_{-0.14}$	\\ 
HBC 374 	& III	& K7$^a$	& 4060	& Optical	& 3.89$^{+0.22}_{-0.23}$	& 0.76$^{+0.04}_{-0.02}$	& 2.58$^{+2.55}_{-1.18}$	& -0.12$^{+0.21}_{-0.19}$	\\ 
LkCa 14 	& III	& M0$^b$	& 3850	& NIR	& 4.26$^{+0.13}_{-0.14}$	& 0.59$^{+0.00}_{-0.01}$	& 8.90$^{+5.01}_{-3.24}$	& -0.67$^{+0.12}_{-0.13}$	\\ 
V827 Tau 	& III	& K7$^a$	& 4060	& Optical	& 3.91$^{+0.33}_{-0.25}$	& 0.77$^{+0.04}_{-0.03}$	& 2.73$^{+5.31}_{-1.33}$	& -0.14$^{+0.23}_{-0.29}$	\\ 
    \hline
     \multicolumn{9}{@{}l@{}}{\hbox to 0pt{\parbox{150mm}{\footnotesize
     \par\noindent
     \footnotemark[$*$] Object classification. II, TD, and III indicate Class II, transitional disk object, and Class III, respectively. Values are from \citet{Kenyon1995} and \citet{Furlan2006}.
     \par\noindent
     \footnotemark[$\dagger$] Spectral types quoted from (a) \citet{Furlan2006} and (b) \citet{Kenyon1995}.
     \par\noindent
     \footnotemark[$\ddagger$] Effective temperatures were determined from the Sp. type - $T_{\mathrm{eff}}$ relation listed in \citet{Kenyon1995}.
     \par\noindent
     \footnotemark[$\S$] The Fe/Na ratio in the optical $I$-band and the Sc/Na ratio in the near-infrared $K$-band were used to estimate surface gravity.
     \par\noindent
     \footnotemark[$\|$] Approximate masses were estimated by comparing the $T_{\mathrm{eff}}$ and the evolutionary tracks of \citet{Siess2000}.
    }\hss}}
   \end{tabular}
 \end{center}
\end{table}

\begin{table}
 \begin{center}
  \caption{Color excesses, the EW (H$\alpha$), and the EW ([O\emissiontype{I}]) of the observed PMS stars.}\label{table-color}
    \begin{tabular}{lccc}
      \hline \hline
      Star 	& $J-K$			& EW H$\alpha$\footnotemark[$*$]	& EW [O\emissiontype{I}] (6300$\AA$) 		\\ 
		& (mag)			& ($\AA$)				& ($\AA$)					\\ \hline
      CY Tau 	& $0.16\pm0.13$		& 70				& 0.49$\pm$0.07\footnotemark[$\dagger$]		\\
      DL Tau 	& $0.71\pm0.23$		& 105				& 0.82$\pm$0.09\footnotemark[$\dagger$]		\\
      DN Tau 	& $0.20\pm0.04$		& 12				& 0.53$\pm$0.08\footnotemark[$\dagger$]		\\
      GI Tau 	& $0.48\pm0.08$		& 19				& 0.48$\pm$0.08\footnotemark[$\dagger$]		\\
      IQ Tau 	& $0.52\pm0.00$		& 7.8				& ---					\\
      LkCa 15 	& $0.36\pm0.09$		& 12.7				& 0.24$\pm$0.04\footnotemark[$\ddagger$]	\\
      V819 Tau 	& $0.00\pm0.00$		& 1.7				& $<$0.05\footnotemark[$\dagger$]		\\
      HBC 374 	& $0.05\pm0.06$		& 3				& $<$0.06\footnotemark[$\dagger$]		\\
      LkCa 14 	& $-0.08\pm0.07$		& 0.9				& ---					\\
      V827 Tau 	& $0.00\pm0.05$		& 1.8				& $<$0.04\footnotemark[$\dagger$]		\\
      \hline
     \multicolumn{3}{@{}l@{}}{\hbox to 0pt{\parbox{85mm}{\footnotesize
     \par\noindent
     \footnotemark[$*$] From \citet{Strom1989}.
     \par\noindent
     \footnotemark[$\dagger$] From \citet{Hartigan1995}.
     \par\noindent
     \footnotemark[$\ddagger$] From \citet{Iguchi2014}.
     }\hss}}
     \end{tabular}
  \end{center}
\end{table}

\clearpage

%%%
% See the manual for the detail.
%%%


\begin{thebibliography}{}
% Journals(e.g. A\&A,ApJ,AJ,NMRAS,PASP ...)
% Authors, Year, Journal, Vol#, Page#
% Journal Title Abbreviation >> http://www.asj.or.jp/pasj/Jabb.html
%\bibitem[Aauthor et al.(2001)]{key-1}
%  Aauthor, A., Bauthor, B., Cauthor, C.\ 2001, PASJ, vol, page
% Books
%\bibitem[Aauthor \& Author(2001a)]{key-2}
%  Aauthor, A., Author, B.\ 2001, Name of Book(Publisher, Tokyo) ch.0
% Books
%\bibitem[Aauthor \& Bauthor(2001b)]{key-3}
%  Aauthor, A., Bauthor, B.\ 2001, Name of Book(Publisher, Tokyo) page0
%......
% Editorial Books
%\bibitem[Dauthor(2001)]{key-n}
%  Dauthor A.~A.\ 2001, in Name of Book,
%   ed.\  D.~Editor (Publisher, Tokyo) page0

\bibitem[Alexander et al.(2006)]{Alexander2006}
 Alexander,~R.~D., Clarke,~C.~J., \& Pringle,~J.~E.\ 2006, \mnras, 369, 229

\bibitem[Appenzeller, Bertout, \& Stahl(2005)]{Appenzeller2005}
 Appenzeller,~I., Bertout,~C., \& Stahl,~O. 2005,\ \aap, 434, 1005

\bibitem[Appenzeller \& Bertout(2013)]{Appenzeller2013}
 Appenzeller,~I., \& Bertout,~C. 2013,\ \aap, 558, 83

\bibitem[Baraffe et al.(1998)]{Baraffe1998}
 Baraffe,~I., Chabrier,~G., Allard,~F., \& Hauschildt,~P.~H.\ 1998, \aap, 337, 403

\bibitem[Bertout \& Genova(2006)]{Bertout2006}
 Bertout,~C., \& Genova,~F.\ 2006, \aap, 460, 499

\bibitem[Bertout, Siess, \& Cabrit(2007)]{Bertout2007}
 Bertout,~C., Siess,~L., \& Cabrit,~S.\ 2007, \aap, 473, L21

\bibitem[Cieza et al.(2005)]{Cieza2005}
 Cieza,~L.~A., Kessler-Silacci,~J.~E.~, Jaffe,~D.~T., Harvey,~P.~M., \& Evans,~II,~N.~J.\ 2005, \apj, 635, 422 

\bibitem[Clarke et al.(2001)]{Clarke2001}
 Clarke,~C.~J., Gendrin,~A., \& Sotomayor,~M.\ 2001, MNRAS, 328, 485

\bibitem[Cohen et al.(1981)]{Cohen1981}
 Cohen,~J.~G., Frogel,~J.~A., Persson,~S.~E., \& Elias,~ J.~H.\ 1981, \apj, 249, 481

\bibitem[Espaillat et al.(2007)]{Espaillat2007}
 Espaillat,~C., Calvet,~N., D'Alessio,~P., Hern\'{a}ndez,~J., Qi,~C., Hartmann,~L., Furlan,~E., \& Watson,~D.~M. 2007,\ \apj, 670, L135

\bibitem[Fischer et al.(2011)]{Fischer2011}
 Fischer,~W., Edwards,~S., Hillenbrand,~L., \& Kwan,~J.\ 2011, \apj, 730, 73

\bibitem[Furlan et al.(2006)]{Furlan2006}
 Furlan,~E. et al.\ 2006, \apjs, 165, 568

\bibitem[Gray(2005)]{Gray2005}
 Gray,~D.~F. \ 2005, in The Observation and Analysis of Stellar Photospheres, (Cambridge University Press)

\bibitem[Haisch, Lada, \& Lada (2001)]{Haisch2001}
 Haisch,~K.~E.,~Jr., Lada,~E.~A., \& Lada,~C.~J.\ 2001, \apj, 553, L153

\bibitem[Hartigan, Edwards, \& Ghandour(1995)]{Hartigan1995}
 Hartigan,~P., Edwards,~S., \& Ghandour,~L.\ 1995, 452, 736

\bibitem[Hashimoto et al.(2012)]{Hashimoto2012}
 Hashimoto,~J. et al.\ 2012, \apjl, 758, L19

\bibitem[Hillenbrand \& White(2004)]{Hillenbrand2004}
 Hillenbrand,~L.~A., \& White,~R.~J.\ 2004, \apj, 604, 741

\bibitem[H{\o}g et al.(2000)]{Hog2000}
 H{\o}g,~E. et al.\ 2000, \aap, 355, L27

\bibitem[Iguchi \& Itoh(2014)]{Iguchi2014}
 Iguchi,~N. \& Itoh,~Y.\ 2014, in prep.

\bibitem[Kenyon \& Hartmann(1995)]{Kenyon1995}
 Kenyon,~S.~J., \& Hartmann,~L.\ 1995, \apjs, 101, 117

\bibitem[Kurosawa, Harries, \& Symington(2006)]{Kurosawa2006}
 Kurosawa,~R., Harries,~T.~J., \& Symington,~N.~H.\ 2006, \mnras, 370, 580

\bibitem[Lejeune \& Schaerer(2001)]{Lejeune2001}
 Lejeune,~T., \& Schaerer,~D.\ 2001, \aap, 366, 538

\bibitem[McClure et al.(2013)]{McClure2013}
 McClure,~M.~K. et al.\ 2013, \apj, 769, 73

\bibitem[Meyer et al.(1997)]{Meyer1997}
 Meyer,~M.~R., Calvet,~N., \& Hillenbrand,~L.~A.\ 1997, \aj, 114, 288

\bibitem[Schiavon, Batalha, \& Barbuy(1995)]{Schiavon1995}
 Schiavon,~R.~P., Batalha,~C., \& Barbuy,~B.\ 1995, \aap, 301, 840

\bibitem[Siess et al.(2000)]{Siess2000}
 Siess,~L., Dufour,~E., \& Forestini,~M.\ 2000, \aap, 358, 593

\bibitem[Slesnick, Carpenter, \& Hillenbrand(2006)]{Slesnick2006}
 Slesnick,~C.~L., Carpenter,~J.~H., \& Hillenbrand,~L.~A.\ 2006, \aj, 131, 3016

\bibitem[Steele \& Jameson(1995)]{Steele1995}
 Steele,~I.~A., \& Jameson,~R.~F.\ 1995, \mnras, 272, 630

\bibitem[Strom et al.(1989)]{Strom1989}
 Strom,~K.~M., Strom,~S.~E., Edwards,~S., Cabrit,~S., \& Skrutskie,~M.~F.\ 1989, \aj, 97, 1451

\bibitem[Takagi et al.(2010)]{Takagi2010}
 Takagi,~Y., Itoh,~Y., \& Oasa,~Y.\ 2010, \pasj, 62, 501

\bibitem[Takagi et al.(2011)]{Takagi2011}
 Takagi,~Y., Itoh,~Y., Oasa,~Y., \& Sugitani,~K.\ 2011, \pasj, 63, 677

\bibitem[Tanaka et al.(2005)]{Tanaka2005}
 Tanaka,~H., Himeno,~Y., \& Ida,~S.\ 2005, \apj, 625, 414

\bibitem[Tanii et al.(2012)]{Tanii2012}
 Tanii,~R. et al.\ 2012, \pasj, 64, 124

\bibitem[van Leeuwen(2007)]{vanLeeuwen2007}
 van Leeuwen,~F.\ 2007, \aap, 474, 653

\bibitem[Zhu et al.(2011)]{Zhu2011}
 Zhu,~Z., Nelson,~R.~P., Hartmann,~L., Espaillat,~C., \& Calvet,~N.\ 2011, \apj, 729, 47


\end{thebibliography}
\end{document}